\documentclass[prl,twocolumn,amsfonts,amssymb,amsmath,floatfix,tightenlines,superscriptaddress]{revtex4-1}

\usepackage{amsfonts,amssymb}
\usepackage{amsmath,times}
\usepackage{graphicx}
\usepackage{hyperref}    
\usepackage{color}
\usepackage{cleveref}
\usepackage[lofdepth,lotdepth,caption=false]{subfig}
\usepackage[normalem]{ulem}

\newcommand{\lead}[1]{\par\noindent\textit{#1.---}}

\begin{document}

\title{Chiral Superconductivity Adjacent to a Wigner Crystal in Rhombohedral Graphene}

\author{Ke Wang}
\email{ke.wang@rochester.edu}
\affiliation{Department of Physics and Astronomy,
University of Rochester, Rochester, New York, USA}
\affiliation{Department of Physics and James Franck Institute,
University of Chicago, Chicago, Illinois, USA}
\affiliation{Leinweber Institute for Theoretical Physics,
University of Chicago, Chicago, Illinois, USA}

\begin{abstract}
It was experimentally observed that Wigner-crystal and chiral superconducting phases occur adjacent to each other in rhombohedral graphene. In this Letter, we explore electronic correlations underlying the two phases and identify the pairing mechanism of chiral superconductivity. Exchange correlations play an essential role in both phases. First, we use self-consistent Hartree-Fock calculations to identify a Wigner-crystal state. Second, to investigate the pairing mechanism, we go beyond charge screening and construct a particle–hole-renormalized interaction through the coupled parquet and Bethe–Salpeter equations. This interaction incorporates exchange correlations while preserving crossing symmetry.  Within this framework, we find that electronic correlations in the quarter metal are strongly enhanced near the crystal phase boundary, generating an effective attraction that favors superconducting pairing. Consequently, chiral superconductivity (CSC) emerges in a narrow sliver adjacent to the crystal. Furthermore, we identify the winding number of the superconducting order parameter as $-3$, indicating its $f$-wave nature. These results connect the emergence of chiral pairing and its location in the phase diagram to the electronic correlations favoring crystallization.
\end{abstract}

\maketitle

\lead{Introduction}
Rhombohedral multilayer graphene has emerged as a natural
platform for exploring strongly correlated electronic
phases. Its stacking-induced, weakly dispersing low-energy
bands enhance the Coulomb interactions
without a moir\'e potential
\cite{Zhou2021Metals,Han2024Correlated}.  The correlation leads to distinct many-body regimes within a single
material system~\cite{Zhou2021,Zhou2022,Chatterjee2022,DelaBarrera2022,Seiler2022,Pantaleon2023,Lu2024,Guo2025,Holleis2025,Han2025,Patterson2025,Yang2025,Auerbach2025,Seo2026,Kalantre2026,Dutta2026}. 

Recent transport measurements have provided evidence for
Wigner crystallization in rhombohedral multilayer graphene
at zero magnetic field \cite{Han2026Wigner,Zhou2026Wigner}. Remarkably, this state borders the
low-density side of a superconducting region, placing
electron crystallization in direct proximity to Cooper
pairing.
Superconductivity in tetralayer and pentalayer graphene
develops within a fully isospin-polarized metal and exhibits signatures consistent with chiral order~\cite{Han2025}.

A central question raised by these observations is the
microscopic origin of superconductivity~\cite{Fradkin2015,Chou2021,Ghazaryan2021,You2022,Cea2022,Qin2023,Dong2023,Li2023,Shavit2023,WangNC,Vituri2025,Dong2026,RevModPhys.84.1383,pichler2026,fcdc-9lm3,Dong2023b,Wang2025,Raines2026,Dong2023a,7xj7-hn98,fdz1-dbf6,sourounis2026}: what
drives Cooper pairing? The proximity to a Wigner crystal (WC)~\cite{Han2026Wigner,Zhou2026Wigner,dong2026crystalscaughtdopingmetallic,ge2026nematicwigner} motivates
a closely related question concerning the relationship
between pairing and electron crystallization. Are these phases
merely competing manifestations of strong electronic
correlations, or can the correlations favoring
crystallization also promote pairing in the neighboring
metal? This possibility suggests investigating whether
electronic fluctuations associated with incipient
crystalline order can generate an effective attractive
interaction before static charge order develops.

In this work, we identify a Coulomb-driven pairing
mechanism associated with incipient Wigner crystallization
in rhombohedral tetralayer graphene. Self-consistent
Hartree--Fock (HF) calculations reveal a competing commensurate
triangular crystal with one doped electron per emergent
unit cell, with exchange playing an essential role in
its stabilization. To investigate how the associated
electronic correlations affect pairing in the neighboring
isospin-polarized metal, we construct a
crossing-symmetric interaction vertex using a
particle--hole parquet formulation with coupled
Bethe--Salpeter equations. This approach incorporates
exchange vertex corrections beyond polarization-bubble
screening while preserving the fermionic exchange
relations of the interaction vertex. Near crystallization,
the enhanced particle--hole correlations generate an
effective attraction in the Cooper channel, favoring
chiral $f$-wave superconductivity. The superconducting
state is confined to a narrow metallic region bordering
the crystal, in qualitative agreement with the
superconducting sleeve observed near the insulating
regime. Our results connect both the emergence of chiral
pairing and its location in the phase diagram to the
electronic correlations favoring Wigner crystallization.

\begin{figure}
    \centering
    \includegraphics[width=\linewidth]{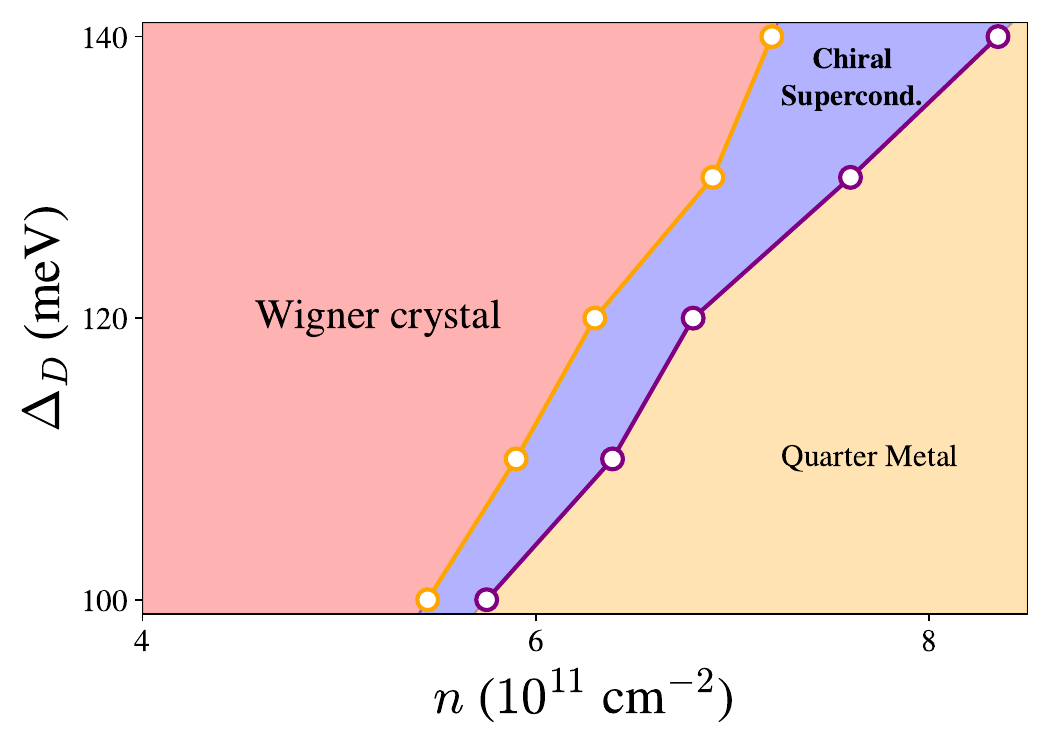}
    \caption{Phase diagram of rhombohedral tetralayer graphene as a
function of electron density $n$ and outer-layer potential
difference $\Delta_D$. The pink, blue, and yellow regions denote
the Wigner crystal, chiral superconducting state, and
spin--valley-polarized quarter metal, respectively.
Superconductivity occupies a narrow sleeve on the metallic
side of crystallization. 
The orange curve marks the estimated crystalline phase boundary,
where the WC order parameter nearly vanishes. The purple curve marks the vanishing of the
superconducting order parameter.
The interaction parameters are $\epsilon_r=15$ and
$d=25\,\mathrm{nm}$.}
    \label{figure1}
\end{figure}

\lead{Model and Hartree--Fock results}
We consider ABCA-stacked tetralayer graphene with
electrons interacting through the gate-screened
Coulomb potential
\begin{equation}
V(\mathbf q)=
\frac{e^2}{2\epsilon_0\epsilon_r|\mathbf q|}
\tanh(d|\mathbf q|),
\label{eq:unified_coulomb}
\end{equation}
where $\epsilon_0$ is the vacuum permittivity,
$\epsilon_r$ is the relative dielectric constant,
and $d$ is the graphene--gate separation.
We use $d=25\,\mathrm{nm}$ and  a large relative dielectric constant $\epsilon_r=15$, following Refs.~\cite{Auerbach2025, dong2026crystalscaughtdopingmetallic}: these constants yielded phase diagrams in good agreement with experimental observations.  We adopt a low-energy non-interacting Hamiltonian\cite{zfmh-rjzc,PhysRevB.82.035409} with eight bands per isospin. We use $\Delta_D$, 
the electrostatic potential difference
between the outermost layers, to model displacement field.  The details are provided in the Supplementary material. 

In the low-electron-density regime, where both CSC and WC emerge experimentally \cite{Han2025} and theoretically (see Fig.~\ref{figure1}), scattering processes that change the electron isospin\cite{PhysRevLett.97.146401} can be neglected . The interaction can therefore be written as
\begin{equation}
\hat V=\frac{1}{2}
\sum_{\mathbf k,\mathbf p,\mathbf q}
\sum_{\alpha,\alpha'}V(\mathbf q)
\hat c^{\dagger}_{\alpha',\mathbf p+\mathbf q}
\hat c^{\dagger}_{\alpha,\mathbf k-\mathbf q}
\hat c_{\alpha,\mathbf k}
\hat c_{\alpha',\mathbf p},
\label{eq:unified_interaction}
\end{equation}
where the internal index $\alpha$
includes spin, valley, layer, and sublattice. In the WC phase, translational symmetry is broken. Thus, we introduce the momentum-off-diagonal density matrix: $
    D^{Q}_{\alpha\beta}(\mathbf k)
=\langle
\hat c^{\dagger}_{\beta,\mathbf k}
\hat c_{\alpha,\mathbf k-\mathbf Q}\rangle.$
This will induce a HF charge amplitude $\Delta^Q_{\alpha\beta}(\mathbf k)=\sum_{\gamma \delta, \mathbf p }\mathcal U^Q_{\alpha\beta,\gamma\delta}(\mathbf k,\mathbf p) D^Q_{ \gamma \delta}(\mathbf p).$ 
Here we introduced the bare antisymmetrized interaction vertex,
\begin{align}
\mathcal U^Q_{\alpha\beta,\gamma\delta}(\mathbf k,\mathbf p)
={}&V(\mathbf Q)\delta_{\alpha\beta}\delta_{\gamma\delta}
-V(\mathbf k-\mathbf p)
\delta_{\alpha\gamma}\delta_{\beta\delta},
\label{3}
\end{align}
which incorporates both direct and exchange terms. This is the irreducible building block for two-particle interactions\cite{Abrikosov1975}.

In the Hartree--Fock calculation, the density matrix is
determined self-consistently at a fixed excess-electron density
$n$. We search for triangular crystalline solutions with
reciprocal lattice vectors $\mathbf Q=m_1\mathbf G_1+m_2\mathbf G_2$,
where $m_1,m_2\in\mathbb Z$,
    $\mathbf G_1=G (\sqrt{3},1)/2$ and
  $
    \mathbf G_2=G(-\sqrt{3},1)/2$.
For a triangular lattice with lattice constant $L$,
$G=4\pi/(\sqrt{3}L)$ and the unit-cell area is
$A_{\mathrm{cell}}=\sqrt{3}L^2/2$.
We primarily consider commensurate WC states with one excess
electron per unit cell, corresponding to a filling
$\nu=A_{\mathrm{cell}}n=1$.
We also examine nearby incommensurate states
to confirm stability of commensurate states.
The calculations use a triangular momentum grid truncated
to a $C_3$-symmetric hexagonal domain with a maximum momentum
$k_{\max}a\simeq 0.13$~\footnote{We checked several larger momentum cutoffs and found only small quantitative changes in the WC solutions.}, where $a$ is the graphene lattice
constant. We sample the first Brillouin zone defined by $\mathbf G_1$
and $\mathbf G_2$ on a $31\times31$ mesh.

\begin{figure}
    \centering
    \includegraphics[width=1.1\linewidth]{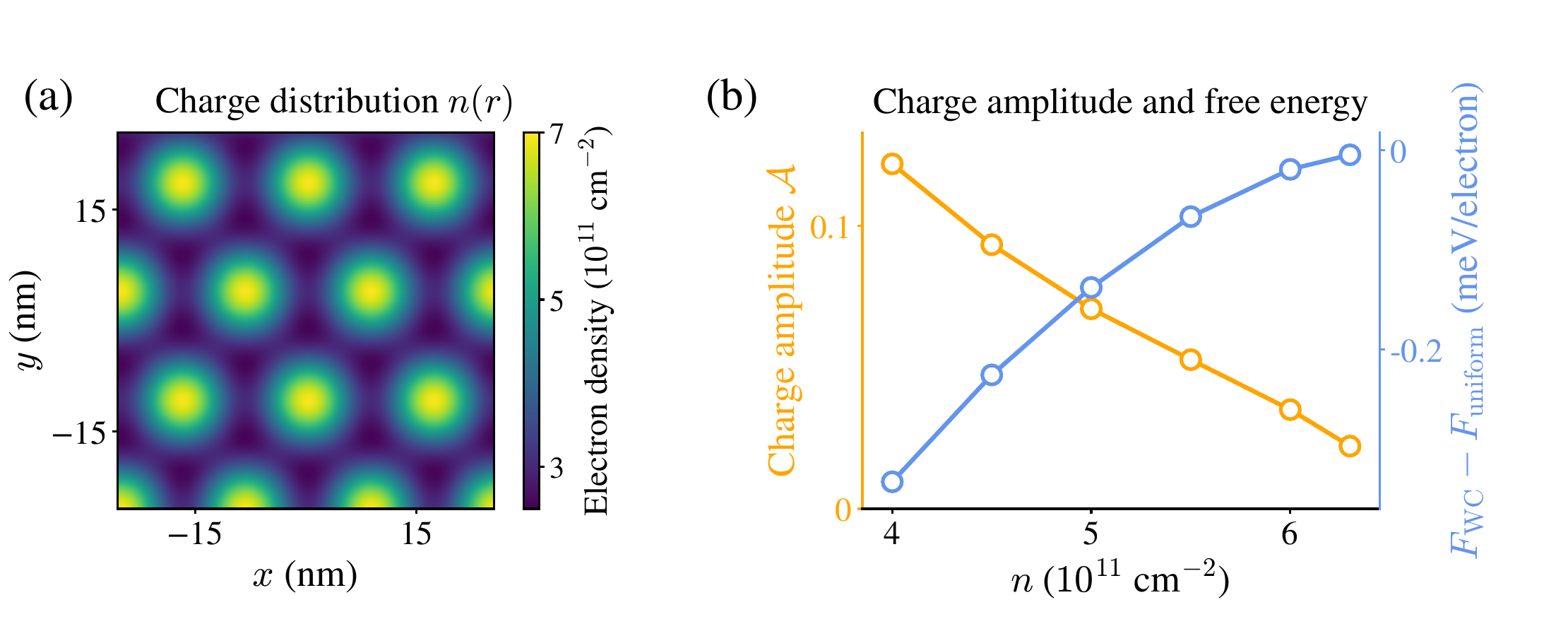}
   \caption{
Hartree--Fock charge order at $\Delta_D=120\,\mathrm{meV}$.
(a) Real-space electron-density distribution of the crystal state at
$n=4\times10^{11}\,\mathrm{cm}^{-2}$.
(b) Charge amplitude $\mathcal{A}$ (orange, left axis) and free-energy
difference $F_{\mathrm{WC}}-F_{\mathrm{uniform}}$ per electron
(blue, right axis) versus density.
All points are $\nu=1$, corresponding to one excess electron
per crystal cell.
}
\label{fig:wc_charge_order}
    \label{figWC}
\end{figure}

We find WC solutions over a substantial region of parameter
space, as shown in Fig.~\ref{figure1}.
For example, at $\Delta_D=120\,\mathrm{meV}$ and
$n=4\times10^{11}\,\mathrm{cm}^{-2}$, the solution exhibits
a clear WC pattern with strong charge-density modulation
and a gap in the HF spectrum.
See Fig.~\ref{figWC}a. To quantify the charge modulation, we expand the electron
density as
$n(\mathbf r)=\sum_{\mathbf G}n_{\mathbf G}
e^{i\mathbf G\cdot\mathbf r}$
and define the first-shell ($f$) charge-modulation amplitude as $
    \mathcal A
    = A_{\mathrm{cell}}
      \max_{\mathbf G \in f }|n_{\mathbf G}|.
$
This amplitude decreases as the crystal--metal boundary
is approached from the crystalline side. See Fig.~\ref{figWC}b.
We use $\mathcal A\simeq 0.02$ as an operational criterion
to estimate the boundary between the WC and the noncrystalline
phase, indicated by the orange curve in Fig.~\ref{figure1}.

Within our numerical resolution, we do not resolve a clear
discontinuity at the transition between the WC and the
neighboring noncrystalline metal. Both the charge
amplitude $\mathcal A$ and the free-energy difference
between the two states approach zero near the phase boundary. See Fig.~\ref{figWC}b.

The direct term
in Eq.~\eqref{3} gives an energy-density difference between
the WC and the noncrystalline metal of
$
    \Delta\mathcal E_H
    =  \sum_{\mathbf G\ne0}V(G)|n_{\mathbf G}|^2 /2.
$
For the repulsive interaction considered here, this contribution
is strictly positive whenever the charge density is modulated.
The Fock term, by contrast, contributes a negative exchange
energy and can favor crystallization. Therefore, exchange correlations play an
essential role in stabilizing WC phases.

\lead{Renormalization of Interaction} Electron scattering amplitudes are not simply described by the bare Coulomb interaction. Rather, they are strongly renormalized by the virtual process of creating particle-hole pairs~\cite{RevModPhys.66.129}. For rhombohedral graphene, the renormalization effects become significantly pronounced due to the nearly flat band.

Two classes of diagrams arise from the direct and exchange
terms in Eq.~\eqref{3}: direct-channel RPA diagrams and
exchange-channel particle--hole ladders, respectively.
Both classes are included and resummed separately
in studies of $^3$He~\cite{PhysRev.169.417} and
spin-fluctuation-mediated superconductivity~\cite{RevModPhys.84.1383}.

 In recent studies of chiral superconductivity in rhombohedral
graphene, the RPA was used to obtain chiral $p$-wave
pairing~\cite{k8s3-dgfs,zfmh-rjzc,fcdc-9lm3}.
We will provide a more quantitative discussion of superconductivity
within the RPA later. Here, we show that ladder diagrams must
be considered on an equal footing with RPA diagrams.
These ladders arise from the exchange terms that
are essential for stabilizing the nearby WC. 

First, RPA generates strong screening of the Coulomb interaction
due to the nearly flat band and the presence of a Fermi
surface~\cite{Mahan2000}. This screening leads to
$V_{\mathrm{RPA}}\ll V$ for $ka\ll 1$.
Thus, RPA strongly {\it weakens} the interaction.
In contrast, particle--hole ladders have the opposite effect.
We find that the solutions of the Bethe--Salpeter equation
for these ladders lie close to a particle--hole singularity.
Consequently, summing {\it only} the particle--hole ladders
yields a strongly enhanced interaction,
$V_{\mathrm{ladder}}\gg V$.

The strongly opposing effects of RPA and ladder resummations
indicate that neither approximation alone provides a complete
description and that both classes of diagrams should be
incorporated in a more consistent scheme. 

Recall that the direct and exchange terms in Eq.~\eqref{3}
arise from antisymmetrization of the interaction, ensuring
that the bare interaction respects Fermi statistics.
To treat these two contributions on an equal footing,
we require the renormalized interaction to obey the same
statistics. To make this requirement explicit,
we write a general quartic interaction: $
\hat\Gamma=\sum\Gamma(12;34)
\hat c^\dagger(1)\hat c(2)\hat c^\dagger(4)\hat c(3)/4.$
Here, each label is shorthand for all microscopic
indices.
Fermi statistics require
\begin{equation}
\begin{aligned}
\Gamma(13;24)&=-\Gamma(12;34),\quad
\Gamma(42;31)=-\Gamma(12;34).
\end{aligned} \label{5}
\end{equation}
These antisymmetry relations are commonly referred to as
crossing symmetry. Neither RPA nor the particle--hole ladder
resummation alone obey the symmetry.
 
\begin{figure}
    \centering
    \includegraphics[width=\linewidth]{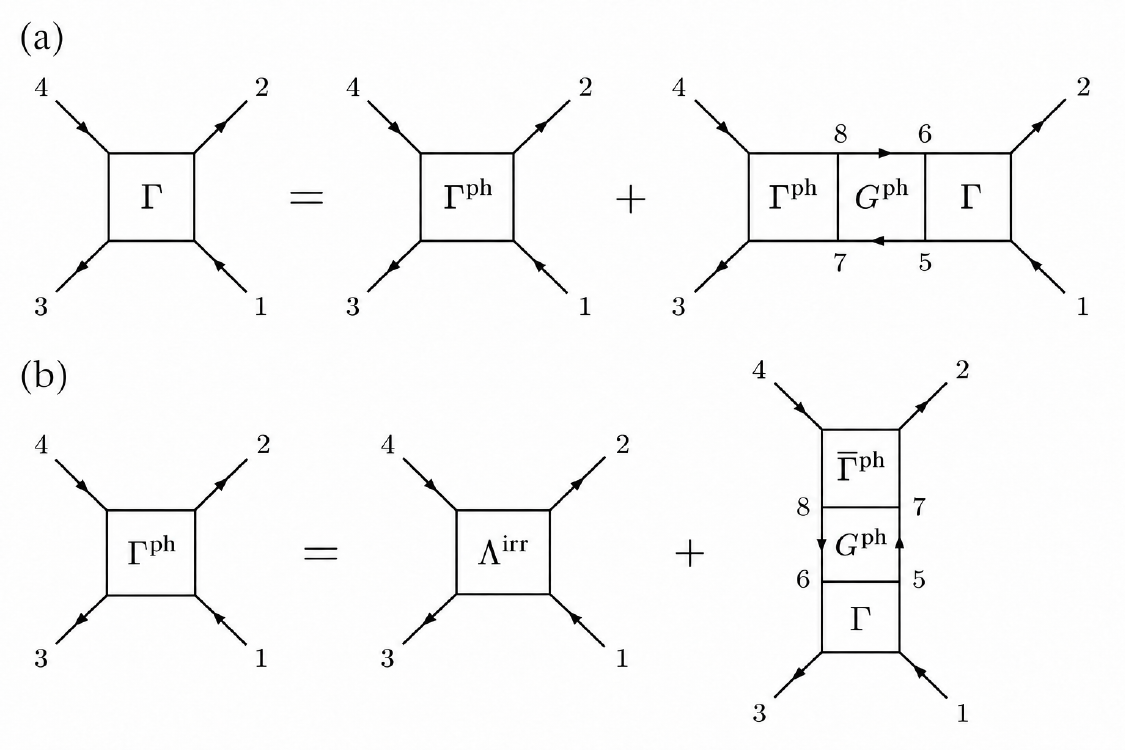}
  \caption{Coupled diagrammatic equations for the four-point vertex
in the particle--hole channels.
(a) Bethe--Salpeter equation for the full vertex $\Gamma$, expressed
in terms of the horizontally irreducible vertex $\Gamma^{\mathrm{ph}}$
and the particle--hole propagator $G^{\mathrm{ph}}$.
(b) Self-consistent construction of $\Gamma^{\mathrm{ph}}$ from
$\Lambda^{\mathrm{irr}}$, which is irreducible in both particle--hole
channels, and the contribution reducible in the vertical
channel. Here $\bar{\Gamma}^{\mathrm{ph}}$ denotes the vertically
irreducible vertex.}
    \label{Feynman}
\end{figure}

This symmetry can be enforced within a {\it parquet}
formalism~\cite{Bickers2004}, which includes particle-particle and two particle-hole  channels. Our goal is to construct the
interaction entering the subsequent pairing calculation.
We therefore retain renormalization in particle--hole
channels but omit particle--particle contributions. The diagrammatic representation of the parquet formalism is shown in Fig.~\ref{Feynman}, and the interaction vertex $\Gamma$ is determined by
\begin{align}
\Gamma^{\mathrm{ph}}(12;34)
&=
\Lambda^{\mathrm{irr}}(12;34)-\Phi(13;24),
\label{6}
\\
\Gamma(12;34)
&=
\Lambda^{\mathrm{irr}}(12;34)
-\Phi(13;24)+\Phi(12;34).
\label{7}
\end{align}
Here, $\Gamma^{\mathrm{ph}}$ is irreducible in the horizontal
particle--hole channel, whereas $\Lambda^{\mathrm{irr}}$
is irreducible in both horizontal and vertical channels.
Horizontal (vertical) irreducibility means that a vertex
cannot be separated into two parts by cutting a pair of
particle--hole propagators in the horizontal (vertical)
channel. The reducible contribution is given by
$
\Phi
=
\Gamma^{\mathrm{ph}}
\left[\mathbf{1}-G^{\mathrm{ph}}\Gamma^{\mathrm{ph}}\right]^{-1}
G^{\mathrm{ph}}\Gamma^{\mathrm{ph}}$, where $G^{\mathrm{ph}}$ denotes a product of particle and
hole propagators.
Provided that $\Lambda^{\mathrm{irr}}$ respects crossing symmetry,
Eq.~\eqref{7} ensures that $\Gamma$ satisfies it too.   See
Ref.~\cite{Bickers2004} and the Supplementary Material for more details.

The results in Eqs.~\eqref{6} and~\eqref{7} are exact at
this stage; approximations enter through the choice of
$\Lambda^{\mathrm{irr}}$. The simplest closure approximates
$\Lambda^{\mathrm{irr}}$ by the antisymmetrized Coulomb
interaction $U$ in Eq.~\eqref{3}. This approximation
was considered sufficiently accurate for liquid
$^3$He~\cite{Anderson1973,PhysRevB.36.156}.
Higher-order corrections can include envelope
diagrams~\cite{Abrikosov1975,Kugler_2018}.

\begin{figure}
    \centering
    \includegraphics[width=1.1\linewidth]{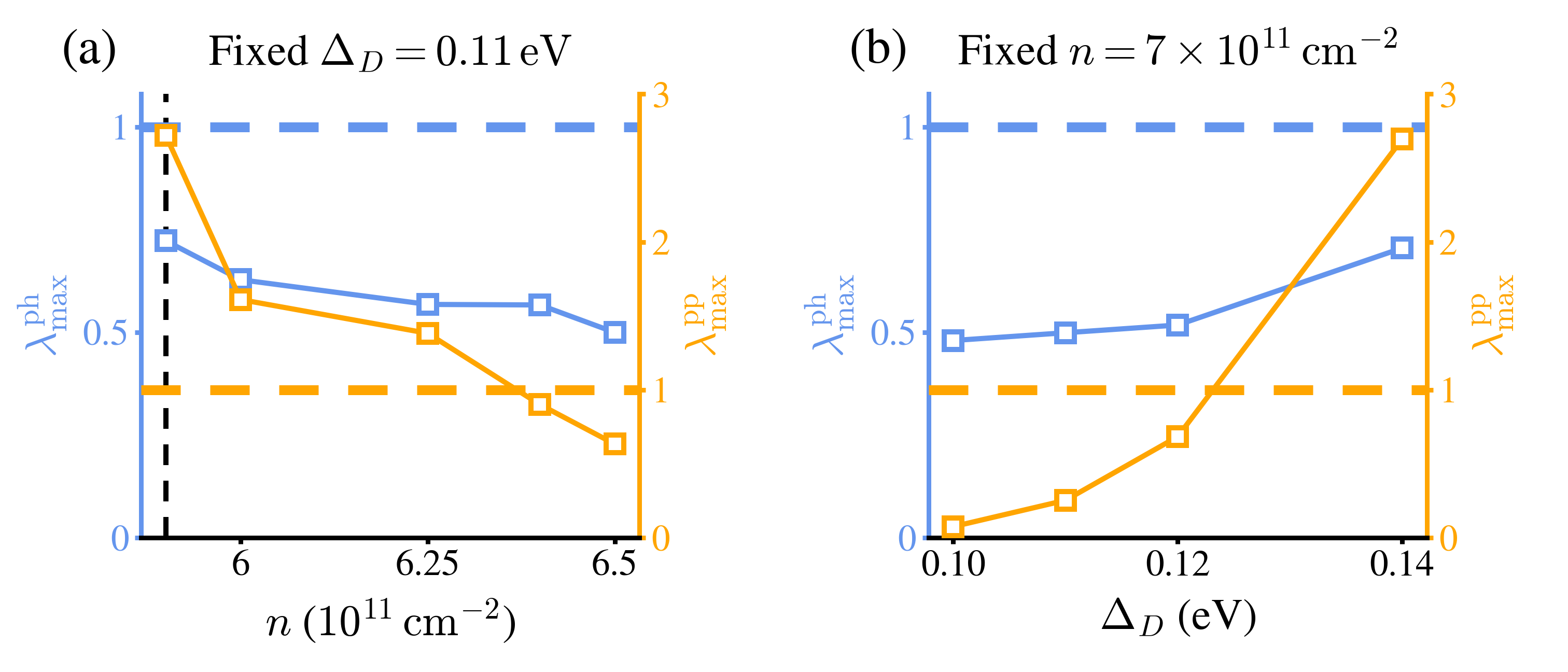}
    \caption{
Maximum particle--hole eigenvalue $\lambda_{\max}^{\mathrm{ph}}$
(blue, left axis) and particle--particle eigenvalue
$\lambda_{\max}^{\mathrm{pp}}$ (orange, right axis), shown versus
(a) carrier density $n$ at fixed $\Delta_D=0.11\,\mathrm{eV}$ and
(b) $\Delta_D$ at fixed $n=7\times10^{11}\,\mathrm{cm}^{-2}$.
Both panels use a gate distance of $25\,\mathrm{nm}$ and
$\epsilon_r=15$. Colored dashed lines mark $\lambda=1$
on the corresponding axes. 
The vertical dashed line marks the phase boundary of the WC.
}
    \label{figure3}
\end{figure}

\begin{figure*}
    \centering
    \includegraphics[width=0.8\linewidth]{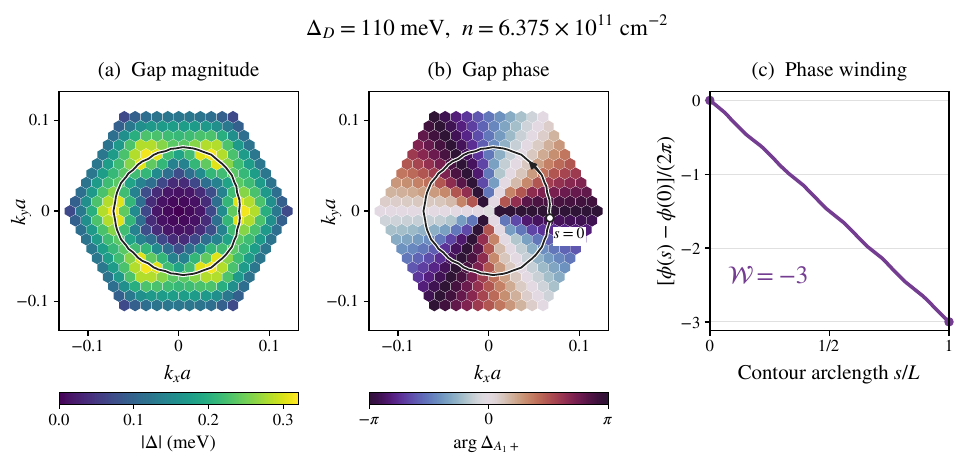}
    \caption{
Superconducting order at $\Delta_D=110$ meV and
$n=6.375\times10^{11}\,\mathrm{cm}^{-2}$, with relative dielectric constant $\epsilon_r=15$ and gate distance $d=25\,\mathrm{nm}$.
The nonlinear gap is evaluated at the temperature $k_{\mathrm B}T=0.01\,\mathrm{meV}$.
(a) Active-band gap magnitude $|\Delta(\mathbf{k})|$, with
a maximum of $0.314$ meV.
(b) Phase of the active-band gap in the $A_1$-positive Bloch
gauge ($A_1$ is the first component of Bloch vector). Each colored hexagon is centered on a sampled momentum;
gray marks the undefined phase at $\mathbf{k}=0$.
The outlined curves in (a) and (b) indicate the normal-state
Fermi contour.
(c) Accumulated phase change along this contour, traversed
counterclockwise from the marked starting point, versus
normalized arclength $s/L$, where $L$ is the contour length.
The total phase change of $-6\pi$ gives winding
$\mathcal{W}=-3$, consistent with a chiral $f$-wave-like
phase texture in the specified gauge.
The contour phase is obtained by piecewise-linear interpolation
of the complex gap on the original momentum mesh.
}
\label{figure4} 
\end{figure*}

\lead{Parquet Vertex and Correlation Effects} 
We adopt a static parquet approximation and project the
parquet equations onto the active band (the only partially
occupied band). Remote-band screening is incorporated
through the constrained-RPA interaction
$V_c(\mathbf k)$~\cite{Aryasetiawan2004}, computed by
excluding only particle--hole transitions entirely within
the active band. We construct $\Lambda^{\mathrm{irr}}$
by antisymmetrizing $V_c(\mathbf k)$ . We solve the coupled equations on the same discrete momentum mesh as in the WC
calculation, with 331 points and a cutoff $k_{\max}a=0.12$.

We now analyze the calculated vertices. From the expression of $\Phi$, one observes the inverse of $ \mathbf{1}-\mathcal{K}\equiv \mathbf{1}-~G^{\mathrm{ph}}\Gamma^{\mathrm{ph}}$. This is in fact the kernel of Bethe-Salpeter (BS) equation $\Gamma(\mathbf{1}-\mathcal{K})=\Gamma^{\mathrm{ph}}$. In the current approximation, the kernel depends only on momenta and may be written as
$\mathcal{K}^{Q}(\mathbf k,\mathbf p)\equiv \mathcal{K}(\mathbf k  - \mathbf Q,\mathbf k; \mathbf p - \mathbf Q, \mathbf p   )$. Here we follow the momentum convention of Eq.~\eqref{3}, and
$\mathbf Q$ is the conserved momentum transfer, equal to
the momentum difference within each particle--hole pair.
At fixed $\mathbf Q$, $\mathcal{K}^{Q}$ can therefore be
regarded as a matrix in momentum space.

A simple but useful diagnostic is the eigensystem of
$\mathcal{K}^{Q}$. Its eigenvalues characterize the
enhancement of the effective interaction in the
corresponding particle--hole channels, while its
eigenvectors describe the momentum structure of these
modes. In particular, an eigenvalue approaching unity
signals proximity to a Bethe--Salpeter pole.
We focus on the largest algebraic eigenvalue
$\lambda^{\mathrm{max}}_{\mathrm{ph}}$, maximized over all sampled
momentum transfers $\mathbf Q$.

We observe a clear trend in the particle--hole channel.
At fixed $\Delta_D$, as the carrier density $n$ is reduced
within the metallic phase toward the WC phase boundary,
the particle--hole eigenvalue
$\lambda^{\mathrm{max}}_{\mathrm{ph}}$ increases,
as shown in Fig.~\ref{figure3}a.
However, it remains below unity, indicating that no particle--hole pole is reached. Within the present approximation and the parameter regime shown in Fig.~\ref{figure1}, we find no evidence
of particle--hole quantum criticality
~\cite{Dong2023a,7xj7-hn98}.

At relatively high densities, such as
$n=7\times 10^{11}\,\mathrm{cm}^{-2}$ in
Fig.~\ref{figure3}(b),
$\lambda^{\mathrm{max}}_{\mathrm{ph}}$ also increases
with increasing $\Delta_D$, consistent with enhanced
correlation effects at larger displacement fields.
Our current calculations are performed at
$\Delta_D\leq 140\,\mathrm{meV}$.
Although this range can be extended, metallic
Wigner-crystal phases may emerge at much
larger displacement fields~\cite{dong2026crystalscaughtdopingmetallic}, a regime beyond the scope
of the present work.

 \lead{Chiral superconductivity}We use the renormalized interaction vertex to investigate pairing in the non-crystalline metal. Define
the vertex
$\Gamma_C(\mathbf{k},\mathbf{k}')
=\Gamma(\mathbf{k},\mathbf{k}';-\mathbf{k}',-\mathbf{k})$. All momenta are measured relative to the selected valley
momentum $\mathbf K_\xi$~\cite{k8s3-dgfs,WangNC}. For a single
isospin flavor, Fermi statistics require
$\Delta(-\mathbf{k})=-\Delta(\mathbf{k})$.
The linearized gap equation is
\begin{equation}
\lambda\Delta(\mathbf{k})
=-\frac{1}{2}\int_{\mathbf{k}'}
\Gamma_C(\mathbf{k},\mathbf{k}')
\frac{1-n_F(\xi_{\mathbf{k}'})-n_F(\xi_{-\mathbf{k}'})}
{\xi_{\mathbf{k}'}+\xi_{-\mathbf{k}'}}
\Delta(\mathbf{k}')
\label{linear}.
\end{equation}
Here $\xi_{\mathbf{k}}=\varepsilon_{\mathbf{k}}-\mu$
denotes the active-band dispersion obtained from $H_0$,
measured relative to the chemical potential $\mu$, and
$\int_{\mathbf{k}}\equiv\int d^2k/(2\pi)^2$.
A leading eigenvalue $\lambda^{\mathrm{pp}}_{\max}>1$ signals a pairing
instability of the normal state. The calculated eigenvalues along cuts
at fixed $\Delta_D$ and fixed $n$ are shown in Fig.~\ref{figure3}. We subsequently solve the nonlinear gap and number
equations, adjusting the chemical potential to preserve the active-band
population.

The resulting interaction supports a finite chiral
superconducting order parameter, as illustrated in
Fig.~\ref{figure4}. As the density increases away from
the WC boundary, the superconducting order
parameter decreases and eventually vanishes.
This defines the phase boundary between the chiral
superconducting state and the metal.
We locate this boundary with a density resolution of
$\Delta n=0.1\times10^{11}\,\mathrm{cm}^{-2}$. We approximate the CSC--WC phase boundary using a small charge-order amplitude, $\mathcal{A}\simeq 0.02$, as a rough criterion. A precise determination requires a comparison of free energies of the two states~\cite{wang2026diff}.

The parent metallic state of the superconducting phase shown in Fig.~\ref{figure1} has a single circular Fermi contour (see Fig.~\ref{figure4}a). At $\Delta_D=0.14\,\mathrm{eV}$, the density range in which
the parent metal has an annular Fermi surface lies
within the WC region. 
Therefore, the FS geometry for the superconducting order is not special. The essential ingredient is the enhanced interaction due to the proximity to WC.

To characterize the phase structure, we fix the arbitrary phase of each
active-band Bloch eigenvector $u(\mathbf{k})$ by requiring its first
component to be real and positive. In this gauge, the superconducting
order parameter has a phase winding of $-3$ around the normal-state
Fermi contour, as shown in Fig.~\ref{figure4}(c).
The same winding is found at the other sampled points within the
superconducting region in Fig.~\ref{figure1}. This phase winding indicates the chiral $f$-wave nature of the superconducting order.

For the parameters of Fig.~\ref{figure4}, we examine several leading
eigenmodes of Eq.~\ref{linear}. The leading mode is $f$-wave,
whereas the next-largest eigenvalue corresponds to a predominantly
$p$-wave mode and lies well below unity. Thus, the $f$-wave-like
channel develops a superconducting instability, while the
$p$-wave-like channel remains below the instability threshold.

We also check pairs $(\mathbf k,\mathbf q-\mathbf k)$
with $\mathbf q\ne0$ at several representative parameter
points, where $\mathbf q$ is the additional pair momentum
relative to $2\mathbf K_\xi$. Near zero temperature, we find that
the $\mathbf q=0$ state has lower energy than the
finite-$\mathbf q$ states.

We next examine RPA-mediated superconductivity. We use the same model parameters, including $\epsilon_r=15$, and include screening contributions from all 32 bands, retaining all flavor-conserving particle--hole transitions. We find no superconducting solution within the parameter regime shown in Fig.~\ref{figure1}. Specifically, at fixed $\Delta_D=0.1375\,\mathrm{eV}$, we scan densities from $3\times10^{11}$ to $8\times10^{11}\,\mathrm{cm}^{-2}$ in increments of $0.1\times10^{11}\,\mathrm{cm}^{-2}$ and find no RPA-induced superconducting instability at $k_BT=0.01\,\mathrm{meV}$. The calculations use the same type of momentum mesh, with cutoff $k_{\max}a=0.15$ and 1951 momentum points.\footnote{At $n=6.5\times10^{11}\,\mathrm{cm}^{-2}$, increasing the number of momentum points to 7651 at the same cutoff also yields no superconducting solution.}

Recent work has investigated Cooper pairing within
a pre-existing magnetic WC background, mediated by
the crystal's collective modes
\cite{miao2026cooperinstabilitymagneticwigner}.
In contrast, we investigate pairing in the neighboring
noncrystalline metal.

\lead{Concluding remarks}
Combining self-consistent Hartree--Fock calculations
with a crossing-symmetric particle--hole parquet
treatment, we identify a Coulomb-driven route to chiral
$f$-wave pairing in rhombohedral tetralayer graphene.
The superconducting state occupies a narrow
isospin-polarized metallic region adjacent to the
crystal, qualitatively consistent with the observed
superconducting sleeve. Our results link chiral pairing
and its location in the phase diagram to exchange-enhanced
correlations near crystallization.

An important next step is a quantitative determination
of the superconducting transition temperature $T_c$.
This requires a self-consistent treatment of fermionic
self-energy corrections and their feedback on the
renormalized pairing interaction, together with the
phase fluctuations intrinsic to two dimensions.
In particular, the superfluid stiffness~\cite{wang2026electrodynamicsignaturesbogoliubovfermi,Wang2026} and the
Berezinskii--Kosterlitz--Thouless transition must be
examined to distinguish the pairing scale from the
temperature at which phase coherence develops.
Such an extension would connect the microscopic
mechanism identified here to the finite-temperature
superconducting behavior.

\lead{Acknowledgment} 
K.W. thanks Kathryn Levin, Chaitanya Murthy, Haoxin Zhou, Yahui Zhang, Hui Yang and Yaar Vituri for helpful discussions. This work was supported by the University of Rochester. K.W. acknowledges support from the Eberly Senior Postdoctoral Fellowship in Quantum Science.

\bibliography{ref}

\end{document}


\title{
       Supplemental Material for  
      Chiral Superconductivity Adjacent to a Wigner Crystal in Rhombohedral Graphene }

\author{Ke Wang}
\email{ke.wang@rochester.edu}
\affiliation{Department of Physics and Astronomy, University of Rochester, Rochester, New York, USA}
\affiliation{Department of Physics and James Franck Institute, University of Chicago, Chicago, Illinois, USA}
\affiliation{Leinweber Institute for Theoretical Physics, University of Chicago, Chicago, Illinois, USA}

     \maketitle
\tableofcontents
\setcounter{equation}{0}
\setcounter{figure}{0}
\setcounter{table}{0}
\setcounter{page}{1}
\makeatletter
\renewcommand{\thesection}{S\arabic{section}}
\renewcommand{\theequation}{S\arabic{equation}}
\renewcommand{\thefigure}{S\arabic{figure}}
\renewcommand{\bibnumfmt}[1]{[S#1]}

\section{Model}

In this section, we briefly review the non-interacting Hamiltonian and the model parameters used in our calculations. We consider rhombohedral tetralayer graphene with $ABCA$ stacking. Each layer contains two sublattices, $A$ and $B$. In the convention adopted here, adjacent layers are arranged such that the $A$ sites in the upper layer lie directly above the $B$ sites in the lower layer.

In the sublattice basis
$
    (A_1,B_1,A_2,B_2,A_3,B_3,A_4,B_4),
$
the continuum Hamiltonian is
\begin{equation}
\mathcal{H}
=
\left(
\begin{array}{cccccccc}
\Delta_D/2+\delta_d & v_0\pi^\dagger & v_4\pi^\dagger & v_3\pi & 0 & \gamma_2/2 & 0 & 0 \\
v_0\pi & \Delta_D/2 & \gamma_1 & v_4\pi^\dagger & 0 & 0 & 0 & 0 \\
v_4\pi & \gamma_1 & \Delta_D/6 & v_0\pi^\dagger & v_4\pi^\dagger & v_3\pi & 0 & \gamma_2/2 \\
v_3\pi^\dagger & v_4\pi & v_0\pi & \Delta_D/6 & \gamma_1 & v_4\pi^\dagger & 0 & 0 \\
0 & 0 & v_4\pi & \gamma_1 & -\Delta_D/6 & v_0\pi^\dagger & v_4\pi^\dagger & v_3\pi \\
\gamma_2/2 & 0 & v_3\pi^\dagger & v_4\pi & v_0\pi & -\Delta_D/6 & \gamma_1 & v_4\pi^\dagger \\
0 & 0 & 0 & 0 & v_4\pi & \gamma_1 & -\Delta_D/2 & v_0\pi^\dagger \\
0 & 0 & \gamma_2/2 & 0 & v_3\pi^\dagger & v_4\pi & v_0\pi & -\Delta_D/2+\delta_d
\end{array}
\right).
\end{equation}
Here,
\begin{equation}
    \pi = \xi k_x + i k_y,
    \qquad
    v_i = \frac{\sqrt{3}}{2}a\gamma_i,
    \quad i=0,3,4,
\end{equation}
where $\mathbf{k}$ is the wave vector measured from the corresponding valley, \(\xi=\pm1\) labels the two valleys, and \(a=2.46\,\text{\AA}\) is the graphene lattice constant.

The terms proportional to \(v_0\) describe intralayer hopping. The parameter \(\gamma_1\) describes direct interlayer hopping between vertically aligned sites \(B_\ell\) and \(A_{\ell+1}\). The skew interlayer hoppings between adjacent layers are parameterized by \(\gamma_3\) and \(\gamma_4\): \(\gamma_3\) couples \(A_\ell\) to \(B_{\ell+1}\), whereas \(\gamma_4\) couples sites on the same sublattice, \(A_\ell\) to \(A_{\ell+1}\) and \(B_\ell\) to \(B_{\ell+1}\). The parameter \(\gamma_2\) describes hopping between next-nearest layers, coupling \(A_1\) to \(B_3\) and \(A_2\) to \(B_4\).

The factors of \(\pi\) and \(\pi^\dagger\) reflect the in-plane geometry of the hopping processes and the chosen sublattice phase convention. In the matrix elements shown above, the \(v_4\) terms have the same momentum dependence as the corresponding intralayer terms, whereas the \(v_3\) terms have the complex-conjugate dependence. The \(\gamma_1\) and \(\gamma_2\) terms are momentum independent within this continuum approximation.

The parameter \(\Delta_D\) denotes the potential-energy difference between the outermost layers induced by an external perpendicular displacement field. We assume a linear potential profile across the four layers, so that the potential-energy difference between adjacent layers is \(\Delta_D/3\). The onsite-energy correction \(\delta_d\) acts only on the outer-layer nondimer sites \(A_1\) and \(B_4\), which have no vertically aligned nearest neighbor in an adjacent layer. While \(\Delta_D\) is tunable, the remaining parameters are fixed to the values 
\(\gamma_0=3.1\,\mathrm{eV}\), \(\gamma_1=0.38\,\mathrm{eV}\), \(\gamma_2=-0.015\,\mathrm{eV}\), \(\gamma_3=-0.29\,\mathrm{eV}\), \(\gamma_4=-0.141\,\mathrm{eV}\), and \(\delta_d=-0.0105\,\mathrm{eV}\).

The remote hoppings associated with \(\gamma_3\) and \(\gamma_2\) generate trigonal warping of the low-energy bands. Their interference with the dominant chiral hopping produces threefold angular dependence, including terms proportional to \(\cos(3\theta)\), where \(\theta\) is the polar angle of \(\mathbf{k}\). These terms reduce the emergent continuous rotational symmetry of the minimal model to threefold rotational symmetry. Thus, \(C_{3z}\) remains a symmetry of the system, whereas \(C_{2z}T\) is absent for \(ABCA\) stacking.

A finite displacement field can produce a Mexican-hat-like dispersion in the low-energy bands. Near \(k=0\), the low-energy states are predominantly localized on the outer layers, where the electrostatic potentials have the largest magnitudes. At finite momentum, these states hybridize with states on the inner layers, which experience smaller potential magnitudes. This hybridization can lower the conduction-band energy at finite momentum, producing a local maximum at \(k=0\) and minima at finite \(k\). The valence band can exhibit the corresponding inverted structure, with a local minimum at \(k=0\) and maxima at finite momentum. These two structures are generally asymmetric because the full Hamiltonian does not possess exact particle--hole symmetry.

If \(\gamma_2\) is neglected, the sites \(A_1\) and \(B_4\) decouple from the remaining sites at \(k=0\), with energies
\begin{equation}
    E_{A_1} = \frac{\Delta_D}{2}+\delta_d,
    \qquad
    E_{B_4} = -\frac{\Delta_D}{2}+\delta_d.
\end{equation}
The common shift \(\delta_d\) means that these energies are not symmetric about zero. For finite \(\gamma_2\), the outer-layer states remain coupled to inner-layer states even at \(k=0\), and their energies must be obtained from the full Hamiltonian. The displacement field can open a gap near charge neutrality.

\section{Hartree--Fock Approximation}

In this section, we describe the screened Coulomb interaction and the self-consistent Hartree--Fock approximation used to study the interacting system.

We consider a long-range Coulomb interaction screened by metallic gates,
\begin{equation}
    V(\mathbf q)
    =
    \frac{1}{4\pi\epsilon_0\epsilon_r}
    \frac{2\pi e^2}{|\mathbf q|}
    \tanh\!\left(d_G|\mathbf q|\right),
\end{equation}
where \(\epsilon_r\) is the effective relative dielectric constant and \(d_G\) is the gate-screening length. We use \(\epsilon_r=15\) and \(d_G=25\,\mathrm{nm}\). Note that its zero-momentum limit is finite:
$V(0)= {e^2d_G}/2\epsilon_0\epsilon_r 
$

One may write the full Hamiltonian as
\begin{equation}
    \hat H_0
    =
    \sum_{\mathbf k,\alpha,\alpha'}
    \hat c^\dagger_{\alpha,\mathbf k}
    K_{\alpha\alpha'}(\mathbf k)
    \hat c_{\alpha',\mathbf k}+\frac{1}{2\Omega}
    \sum_{\mathbf k,\mathbf p,\mathbf q}
    \sum_{\alpha,\alpha'}
    V(\mathbf q)
    \hat c^\dagger_{\alpha',\mathbf p+\mathbf q}
    \hat c^\dagger_{\alpha,\mathbf k-\mathbf q}
    \hat c_{\alpha,\mathbf k}
    \hat c_{\alpha',\mathbf p}.
\end{equation}
where \(\alpha\) is an index encompassing layer, sublattice, spin, and valley. \(\Omega\) is the physical area of the two-dimensional system.  In the main-text, we set $\Omega=1$ for simplicity.  Accordingly, the momentum sums satisfy
$
    \frac{1}{\Omega}\sum_{\mathbf k}
    \longrightarrow
    \int\frac{d^2k}{(2\pi)^2}
$
in the thermodynamic limit. We neglect the dependence of the interaction on internal indices, including the layer index.
\subsection{Uniform HF}
Here we consider density matrices diagonal in momentum and define the equal-time single-particle density matrix as $
    D_{\alpha\alpha'}(\mathbf k)
    =
    \langle
    \hat c^\dagger_{\alpha',\mathbf k}
    \hat c_{\alpha,\mathbf k}
\rangle. $
The Hartree--Fock decoupling then gives
\begin{equation}
\begin{aligned}
    \hat V_{\mathrm{MF}}
    ={}&
    \sum_{\mathbf k,\alpha,\alpha'}
    \hat c^\dagger_{\alpha,\mathbf k}
    \left[
        H_{\alpha\alpha'}
        +F_{\alpha\alpha'}(\mathbf k)
    \right]
    \hat c_{\alpha',\mathbf k}
    -
    \frac{1}{2}
    \sum_{\mathbf k}
    \operatorname{tr}
    \left[
        \left(H+F(\mathbf k)\right)D(\mathbf k)
    \right],
\end{aligned}
\end{equation}
where the trace is over internal indices. The Hartree and Fock matrices are
\begin{equation}
\begin{aligned}
    H_{\alpha\alpha'}=
    \frac{\delta_{\alpha\alpha'}}{\Omega}
    V(0)
    \sum_{\mathbf p,\beta}
    D_{\beta\beta}(\mathbf p), \quad
    F_{\alpha\alpha'}(\mathbf k)
    =
    -\frac{1}{\Omega}
    \sum_{\mathbf p}
    V(\mathbf k-\mathbf p)
    D_{\alpha\alpha'}(\mathbf p).
\end{aligned}
\end{equation}
The self-consistent single-particle Hamiltonian is therefore $
    h_{\mathrm{HF}}(\mathbf k)
    =
    K(\mathbf k)+H+F(\mathbf k),$
with eigenvectors and eigenvalues determined by $
    h_{\mathrm{HF}}(\mathbf k)\phi_n(\mathbf k)
    =
    \epsilon_n^{\mathrm{HF}}(\mathbf k)\phi_n(\mathbf k).
$
The components of \(\phi_n(\mathbf k)\) are defined through $
    \lvert\psi_{n,\mathbf k}^{\mathrm{HF}}\rangle
    =
    \sum_\alpha
    \phi_n^\alpha(\mathbf k)
    \lvert\alpha,\mathbf k\rangle.
$
The density matrix can then be reconstructed as
\begin{equation}
    D_{\alpha\alpha'}(\mathbf k)
    =
    \sum_n
    f_n(\mathbf k)
    \phi_n^\alpha(\mathbf k)
    \phi_n^{\alpha',*}(\mathbf k),\quad  f_n(\mathbf k)
    =
    \frac{1}{
        \exp\!\left[
            \bigl(\epsilon_n^{\mathrm{HF}}(\mathbf k)-\mu\bigr)
            /(k_{\mathrm B}T)
        \right]+1
    }
\end{equation}
 At zero temperature, this expression reduces to a sum over occupied states at each momentum. Because $H$ and $F$ depend on $D$, these equations must be solved self-consistently. Starting from a trial density matrix, we construct and diagonalize \(h_{\mathrm{HF}}\), determine the chemical potential \(\mu\) at the prescribed electron density, and update \(D\). This procedure is repeated until convergence. In particular, the Fock matrix at a given momentum depends on the density matrix at all momenta through the exchange kernel \(V(\mathbf k-\mathbf p)\).

\subsection{Hartree--Fock treatment of the Wigner-crystal phase}

The Wigner-crystal (WC) phase spontaneously breaks translation symmetry through a periodic modulation of the electron density. To describe this phase, we generalize the density matrix to allow coherence between momenta differing by reciprocal lattice vectors of the WC. For a chosen crystal lattice, we write each momentum as \(\mathbf k+\mathbf G\), where \(\mathbf k\) lies in the fisrt Brillouin zone and \(\mathbf G\) is a WC reciprocal lattice vector. The density matrix becomes
\begin{equation}
    D_{\alpha\alpha'}^{\mathbf G\mathbf G'}(\mathbf k)
    =
    \left\langle
    \hat c^\dagger_{\alpha',\mathbf k+\mathbf G'}
    \hat c_{\alpha,\mathbf k+\mathbf G}
    \right\rangle.
\end{equation}
The components with \(\mathbf G\neq\mathbf G'\) describe coherence between distinct momenta and allow a spatially nonuniform density. In this enlarged basis, the Hartree--Fock Hamiltonian is
\begin{equation}
    [h_{\mathrm{HF}}(\mathbf k)]_{\alpha\alpha'}^{\mathbf G\mathbf G'}
    =
    \delta_{\mathbf G\mathbf G'}
    K_{\alpha\alpha'}(\mathbf k+\mathbf G)
    +
    H_{\alpha\alpha'}^{\mathbf G\mathbf G'}
    +
    F_{\alpha\alpha'}^{\mathbf G\mathbf G'}(\mathbf k).
\end{equation}
The Hartree contribution is determined by the Fourier components of the electron density,
\begin{equation}
    n(\mathbf r)
    =
    \sum_{\mathbf Q}
    n_{\mathbf Q}e^{i\mathbf Q\cdot\mathbf r},
    \quad
    n_{\mathbf Q}
    =
    \frac{1}{\Omega}
    \sum_{\mathbf k\in\mathrm{BZ}}
    \sum_{\mathbf G,\alpha}
    D_{\alpha\alpha}^{\mathbf G+\mathbf Q,\mathbf G}(\mathbf k).
\end{equation}
where \(\mathbf Q\) is a WC reciprocal lattice vector and $n_{\mathbf 0}=n$ is the average electron density. The Hartree and Fock matrices are $H_{\alpha\alpha'}^{\mathbf G\mathbf G'}=
    \delta_{\alpha\alpha'}
    V(\mathbf G-\mathbf G')
    n_{\mathbf G-\mathbf G'},$ and
\begin{equation}
    F_{\alpha\alpha'}^{\mathbf G\mathbf G'}(\mathbf k)
    =
    -\frac{1}{\Omega}
    \sum_{\mathbf p\in\mathrm{BZ}}
    \sum_{\mathbf L}
    V(\mathbf k-\mathbf p+\mathbf G-\mathbf L)
    D_{\alpha\alpha'}^{\mathbf L,\mathbf L+\mathbf G'-\mathbf G}
    (\mathbf p).
\end{equation}
Here \(\mathbf L\) runs over reciprocal lattice vectors. These expressions use the same internal-index-independent interaction as in the homogeneous calculation.

Unlike the homogeneous case, the Hartree potential now contains nonzero reciprocal-lattice harmonics and therefore cannot be absorbed entirely into the chemical potential. Its contribution to the total energy is $
    E_{\mathrm H}
    \sim
    \sum_{\mathbf Q}
    V(\mathbf Q)|n_{\mathbf Q}|^2/2.
$
At fixed average density, only the \(\mathbf Q=0\) contribution is common to all competing states and may be omitted. The terms with \(\mathbf Q\neq0\) depend on the density modulation and must be retained when comparing the WC and homogeneous phases.

For a chosen crystal lattice, the self-consistent calculation starts from a trial density matrix containing a weak periodic modulation. Diagonalizing \(h_{\mathrm{HF}}(\mathbf k)\) gives reconstructed bands and eigenvectors \(\phi_n^{\alpha\mathbf G}(\mathbf k)\), from which the density matrix is updated according to
$$
    D_{\alpha\alpha'}^{\mathbf G\mathbf G'}(\mathbf k)
    =
    \sum_n f_n(\mathbf k)
    \phi_n^{\alpha\mathbf G}(\mathbf k)
    \phi_n^{\alpha'\mathbf G',*}(\mathbf k).
$$
The chemical potential is adjusted to maintain the prescribed density, and the Hartree and Fock matrices are recomputed until self-consistency is reached. At zero temperature, competing solutions are compared using
\begin{equation}
    E_{\mathrm{HF}}
    =
    \sum_{\mathbf k\in\mathrm{BZ}}
    \operatorname{Tr}
    \left[
        \left(
            K_{\mathrm{WC}}(\mathbf k)
            +\frac{1}{2}H
            +\frac{1}{2}F(\mathbf k)
        \right)
        D(\mathbf k)
    \right],
\end{equation}
where
\(
[K_{\mathrm{WC}}(\mathbf k)]_{\alpha\alpha'}^{\mathbf G\mathbf G'}
=
\delta_{\mathbf G\mathbf G'}
K_{\alpha\alpha'}(\mathbf k+\mathbf G)
\),
and the trace includes both internal and reciprocal-lattice indices. Comparing converged solutions obtained from different initial conditions and candidate lattices identifies the energetically preferred state within the variational space considered.

\section{Particle--Hole Renormalization Vertex}
At second order in perturbation theory, particle--hole corrections include both bubble diagrams, as in the random-phase approximation (RPA), and ladder diagrams. Resumming these contributions independently neglects the mixing between the two channels, which becomes explicit at third order. Representative third-order include: (1) \textit{Ladder$*$Ladder}, the third-order ladder diagram; (2) \textit{Loop$*$Loop}, the third-order RPA diagram; (3) \textit{Loop$*$Ladder}, a mixed diagram whose external legs connect to a ladder structure on one side and an RPA structure on the other; (4) \textit{Loop(Ladder)}, an RPA loop dressed by ladder corrections, such as an interaction line inserted within the loop; and (5) \textit{Ladder(Loop)}, a ladder diagram with bubble corrections to its interaction lines.

As the perturbative order increases, the number and variety of diagrams grow, motivating a systematic resummation scheme. Below, we first introduce the interaction vertex arising from Hartree--Fock (HF) linear-response theory, which provides a useful starting point for understanding these corrections. We then present parquet resummation as a systematic framework for treating the coupling between scattering channels. 

\subsection{ HF Linear Response and associated Vertex }

{\it HF Linear Response.--} Consider a one-body perturbation of the form
\begin{equation}
\delta H(Q)=\sum_{p,\alpha\alpha'}
 c^\dagger_{\alpha,\mathbf p-\mathbf Q}
 \Delta_{\alpha\alpha'}(\mathbf p-\mathbf Q,\mathbf p)
 \hat c_{\alpha',\mathbf p}.
\label{147}
\end{equation}
We allow $\Delta$ to depend independently on both $p$ and $Q$. To first order in this perturbation, the off-diagonal Green's function is
\begin{align}
G_{\beta\beta'}(p-Q,p)
&=-\langle c_\beta(p-Q)c^\dagger_{\beta'}(p)\rangle
\nonumber=\left\langle c_\beta(p-Q)c^\dagger_{\beta'}(p)
 c^\dagger_{\alpha,\mathbf p-\mathbf Q}
 \Delta_{\alpha\alpha'}(\mathbf p-\mathbf Q,\mathbf p)
 \hat c_{\alpha',\mathbf p}\right\rangle
\nonumber\\
&=G_{\beta\alpha}(p-Q)
 \Delta_{\alpha\alpha'}(\mathbf p-\mathbf Q,\mathbf p)
 G_{\alpha'\beta'}(p).
\end{align}
Here and below, repeated internal indices are summed unless otherwise specified. We focus on a static perturbation with $Q=(0,\mathbf Q)$ and $\mathbf Q\ne0$. The corresponding equal-time density matrix is
\begin{equation}
D_{\alpha\alpha'}(\mathbf p-\mathbf Q,\mathbf p)
=\langle\hat c^\dagger_{\alpha'}(\mathbf p)
 \hat c_\alpha(\mathbf p-\mathbf Q)\rangle
=T\sum_{p_0}G_{\alpha\alpha'}(p-Q,p).
\end{equation}
Writing the Green's function in its spectral representation, we evaluate the Matsubara sum:
\begin{equation}
G(p)=\sum_n\frac{|\phi_n(\mathbf p)\rangle
 \langle\phi_n(\mathbf p)|}{ip_0-\epsilon_n(\mathbf p)},\quad \frac{1}{\beta}\sum_\omega
 \frac{1}{i\omega-\xi}\frac{1}{i\omega-\xi'}
=\frac{n_F(\xi)-n_F(\xi')}{\xi-\xi'}.
\end{equation}
The contribution from a given pair of bands is therefore
\begin{equation}
L_{n,m}(\mathbf p-\mathbf Q,\mathbf p)
 \phi_n(\beta,\mathbf p-\mathbf Q)
 \phi_n^*(\alpha,\mathbf p-\mathbf Q) \times\Delta_{\alpha\alpha'}(\mathbf p-\mathbf Q,\mathbf p)
 \phi_m(\alpha',\mathbf p)\phi_m^*(\beta',\mathbf p).
\end{equation}
Equivalently, the density matrix can be written as
\begin{align}
D(\mathbf p-\mathbf Q,\mathbf p)
=\sum_{n,m}L_{n,m}(\mathbf p-\mathbf Q,\mathbf p)
 |\phi_n(\mathbf p-\mathbf Q)\rangle
\langle\phi_n(\mathbf p-\mathbf Q)|
 \Delta|\phi_m(\mathbf p)\rangle\langle\phi_m(\mathbf p)|,\label{S20}
\end{align}
where the multiband Lindhard function is defined by
\begin{equation}
L_{n,m}(\mathbf p-\mathbf Q,\mathbf p)
=\frac{n_F(\xi_n(p-Q))-n_F(\xi_m(p))}
 {\xi_n(p-Q)-\xi_m(p)}.
\end{equation}
The induced density matrix, in turn, generates Fock and Hartree contributions:
\begin{align}
\text{Fock:}\quad
&-\sum_{p,q}V(q-p)\hat c^\dagger_{\alpha,\mathbf p-\mathbf Q}
 D_{\alpha\alpha'}(q-Q,q)\hat c_{\alpha',\mathbf p}
 +\mathrm{h.c.},\\
\text{Hartree:}\quad
&V(-Q)\sum_{p,q}D_{\alpha\alpha}(q-Q,q)
 \hat c^\dagger_{\alpha',\mathbf p-\mathbf Q}
 \hat c_{\alpha',\mathbf p}+\mathrm{h.c.}
\end{align}
The induced HF charge amplitude field is thus
\begin{align}
\Delta^{\mathrm{ind}}_{\beta\beta'}(p-Q,p)=\delta_{\beta\beta'}V(-Q)
 \sum_{q,\alpha}D_{\alpha\alpha}(q-Q,q)
-\sum_q V(q-p)D_{\beta\beta'}(q-Q,q),
\end{align}
Now we substitute the linear response of the density matrix from Eq.~\ref{S20} and get
\begin{eqnarray}
&&\Delta^{\mathrm{ind}}_{\beta\beta'}(q-Q,q)
=\delta_{\beta\beta'}V(-Q)\sum_{p,\alpha,n,m}
 L_{n,m}(\mathbf p-\mathbf Q,\mathbf p)
\times\phi_n(\alpha,\mathbf p-\mathbf Q)
 \Delta_{nm}(\mathbf p-\mathbf Q,\mathbf p)
 \phi_m^*(\alpha,\mathbf p)
\nonumber\\
&&-\sum_{p,n,m}V(p-q)L_{n,m}(\mathbf p-\mathbf Q,\mathbf p)
\times\phi_n(\beta,\mathbf p-\mathbf Q)
 \Delta_{nm}(\mathbf p-\mathbf Q,\mathbf p)
 \phi_m^*(\beta',\mathbf p),
\end{eqnarray}
where $\Delta_{nm}(\mathbf p-\mathbf Q,\mathbf p)
=\langle\phi_n(\mathbf p-\mathbf Q)|
 \Delta|\phi_m(\mathbf p)\rangle
=\sum_{\gamma,\gamma'}\phi_n^*(\gamma,\mathbf p-\mathbf Q)
 \Delta_{\gamma\gamma'}\phi_m(\gamma',\mathbf p).$
This relation defines the linear response kernel
\begin{align}
\mathcal K^Q_{\beta\beta',\gamma\gamma'}(\mathbf q,\mathbf p)
&=\sum_{n,m}L_{n,m}(\mathbf p-\mathbf Q,\mathbf p)
 \phi_n^*(\gamma,\mathbf p-\mathbf Q)\phi_m(\gamma',\mathbf p) \label{S25}
\\
&\quad\times\Bigl\{V(-Q)\delta_{\beta\beta'}
 \sum_\alpha\phi_n(\alpha,\mathbf p-\mathbf Q)\phi_m^*(\alpha,\mathbf p)
-V(\mathbf p-\mathbf q)
 \phi_n(\beta,\mathbf p-\mathbf Q)\phi_m^*(\beta',\mathbf p)\Bigr\}.\nonumber
\end{align}
The induced field then takes the compact form
\begin{equation}
\Delta^{\mathrm{ind}}_{\beta\beta'}(\mathbf q-\mathbf Q,\mathbf q)
=\sum_{p,\gamma\gamma'}
 \mathcal K^Q_{\beta\beta',\gamma\gamma'}(\mathbf q,\mathbf p)
 \Delta_{\gamma\gamma'}(\mathbf p-\mathbf Q,\mathbf p).
\end{equation}
The transfer $Q$ is conserved in this response channel because the internal propagators and interactions conserve momentum at each vertex. The Hartree contribution is independent of the outgoing momentum $\mathbf q$ and therefore produces a momentum-independent shift. It must nevertheless be retained for a consistent HF response.

Since the fermions respond to the total field, the self-consistency condition is
\begin{align}
&\Delta^{\mathrm{tot}}_{\beta\beta'}(\mathbf q-\mathbf Q,\mathbf q)
=\Delta^{\mathrm{ext}}_{\beta\beta'}(\mathbf q-\mathbf Q,\mathbf q)
+\sum_{p,\gamma\gamma'}
 \mathcal K^Q_{\beta\beta',\gamma\gamma'}(\mathbf q,\mathbf p)
 \Delta^{\mathrm{tot}}_{\gamma\gamma'}(\mathbf p-\mathbf Q,\mathbf p).\nonumber \\
 \iff  &\sum_{p,\gamma\gamma'}\Bigl[
 \delta_{\beta\gamma}\delta_{\beta'\gamma'}\delta(\mathbf q,\mathbf p)
 -\mathcal K^Q_{\beta\beta',\gamma\gamma'}(\mathbf q,\mathbf p)\Bigr]
 \Delta^{\mathrm{tot}}_{\gamma\gamma'}(\mathbf p-\mathbf Q,\mathbf p)=\Delta^{\mathrm{ext}}_{\beta\beta'}(\mathbf q-\mathbf Q,\mathbf q).
\end{align}
This gives the linear response reponse within HF framework.

{\it Reconstructing the Vertex.--}
The linear response equation relates an external perturbation to the total field generated by the interaction and the fermionic response. The shared kernel identified above suggests a direct construction of the renormalized interaction from this equation.

For fixed $Q$, introduce composite indices $
A=(q,\beta,\beta')$ and $B=(p,\gamma,\gamma')$.
Each index specifies a momentum and a pair of internal atomic indices. The induced field can then be expressed either in terms of the total field or in terms of the density matrix:
\begin{equation}
\Delta_A^{\mathrm{ind}}=\sum_B\mathcal K^Q_{AB}\Delta_B^{\mathrm{tot}},
\quad
\Delta_A^{\mathrm{ind}}=\sum_B\mathcal U^Q_{AB}D_B,\quad U^Q_{\beta\beta',\gamma\gamma'}(\mathbf q,\mathbf p)
=V(Q)\delta_{\beta\beta'}\delta_{\gamma\gamma'}
 -V(q-p)\delta_{\beta\gamma}\delta_{\beta'\gamma'}.
\end{equation}  Now choose an external field of the form
$
\Delta_A^{\mathrm{ext}}=\sum_B\mathcal U^Q_{AB}\eta_B,
$
where $\eta_B$ is an arbitrary prescribed particle--hole amplitude. This choice serves as an algebraic probe: the applied field is obtained by attaching a bare interaction to $\eta$. Define the dressed interaction through $
\Delta_A^{\mathrm{tot}}=\sum_B\Gamma_{AB}\eta_B. $
Substitution into the self-consistency equation gives
\begin{equation}
\Gamma\eta=\mathcal U\eta+\mathcal K\Gamma\eta
\quad\Longrightarrow\quad
(I-\mathcal K)\Gamma=\mathcal U,
\qquad
\Gamma=(I-\mathcal K)^{-1}\mathcal U.
\end{equation}
Here $I$ is the identity in the space of composite particle--hole indices, and $\mathcal K$ is given by Eq.~\eqref{S25}. Restoring the indices, we obtain
\begin{align}
&\sum_{q_1,\gamma\gamma'}\Bigl[
 \delta_{\beta\gamma}\delta_{\beta'\gamma'}\delta(\mathbf q,\mathbf q_1)
 -\mathcal K^Q_{\beta\beta',\gamma\gamma'}(\mathbf q,\mathbf q_1)\Bigr]
 \Gamma^Q_{\gamma\gamma',\alpha\alpha'}(\mathbf q_1,\mathbf p)
=\mathcal U^Q_{\beta\beta',\alpha\alpha'}(\mathbf q,\mathbf p).
\end{align}
The resulting $\Gamma^Q$ is the interaction renormalized in the selected particle--hole channel. This construction is an RPA-type resummation of the antisymmetrized interaction. Its iterative form is
\begin{align}
\Gamma^Q_{\beta\beta',\alpha\alpha'}(\mathbf q,\mathbf p)
&=\mathcal U^Q_{\beta\beta',\alpha\alpha'}(\mathbf q,\mathbf p)
+\sum_{q_1,\gamma\gamma'}
 \mathcal K^Q_{\beta\beta',\gamma\gamma'}(\mathbf q,\mathbf q_1)
 \Gamma^Q_{\gamma\gamma',\alpha\alpha'}(\mathbf q_1,\mathbf p).
\end{align}
To express this equation in the band basis, define
\begin{align}
\Gamma^Q_{nm,ij}(\mathbf q_1,\mathbf p)
&=\phi_n^*(\gamma,\mathbf q_1-\mathbf Q)
 \phi_m(\gamma',\mathbf q_1)
 \Gamma^Q_{\gamma\gamma',\alpha\alpha'}(\mathbf q_1,\mathbf p)
\times\phi_i(\alpha,\mathbf p-\mathbf Q)
 \phi_j^*(\alpha',\mathbf p),\\
\mathcal U^Q_{uv,ij}(\mathbf q,\mathbf p)
&=\phi_u^*(\beta,\mathbf q-\mathbf Q)\phi_v(\beta',\mathbf q)
 \mathcal U^Q_{\beta\beta',\alpha\alpha'}(\mathbf q,\mathbf p)
 \times\phi_i(\alpha,\mathbf p-\mathbf Q)
 \phi_j^*(\alpha',\mathbf p).
\end{align}
Here one sub-index represents one band. The band-basis response kernel is
\begin{align}
\mathcal K^Q_{uv,nm}(\mathbf q,\mathbf q_1)
&=L_{n,m}(\mathbf q_1-\mathbf Q,\mathbf q_1)
 \Bigl\{V(-Q)
 \langle\phi_u(\mathbf q-\mathbf Q)|\phi_v(\mathbf q)\rangle
\times\langle\phi_m(\mathbf q_1)|\phi_n(\mathbf q_1-\mathbf Q)\rangle
\nonumber\\
&\quad-V(\mathbf q_1-\mathbf q)
 \langle\phi_u(\mathbf q-\mathbf Q)|\phi_n(\mathbf q_1-\mathbf Q)\rangle
\times\langle\phi_m(\mathbf q_1)|\phi_v(\mathbf q)\rangle
 \Bigr\}.
\end{align}
The Bethe--Salpeter equation then becomes
\begin{align}
\Gamma^Q_{uv,ij}(\mathbf q,\mathbf p)
&=\mathcal U^Q_{uv,ij}(\mathbf q,\mathbf p)
+\sum_{n,m,q_1}\mathcal K^Q_{uv,nm}(\mathbf q,\mathbf q_1)
 \Gamma^Q_{nm,ij}(\mathbf q_1,\mathbf p).
\end{align}
This corresponds to a resummation in a single (horizontal) particle--hole channel.

\subsection{Parquet Vertex}
Now we develop a parquet diagrammatic renormalization of interactions. This essentially sums over two particle-hole channels, both horizontally and vertically.  We adopt the convention in which the arrow associated with $G(12)$ points from $1$ to $2$. Incoming and outgoing external lines are therefore associated with creation and annihilation operators, respectively. This convention follows the operator assignments in the tree-level vertex and the amputation of external propagators. 

Write the full 4-point vertex as $\Gamma$ and its effective interaction reads $$
\hat\Gamma=\sum\Gamma(12;34)
\hat c^\dagger(1)\hat c(2)\hat c^\dagger(4)\hat c(3)/4.$$
 The ordering of the second particle--hole pair is reversed so that the two pairs can serve as matrix indices of $\Gamma$.
Fermionic antisymmetry gives the crossing relations
\begin{equation}
    \Gamma(42;31)=-\Gamma(12;34),
    \qquad
    \Gamma(13;24)=-\Gamma(12;34).
\end{equation}
Diagrammatically, these relations correspond to exchanging the two creation-operator legs or the two annihilation-operator legs. 

To formulate the equation for the full vertex, we first distinguish the two particle--hole channels according to the orientation of their internal propagators in the diagrams:
\begin{itemize}
    \item \emph{Horizontal irreducibility.} A diagram is irreducible in the horizontal channel if a vertical cut through two horizontal particle--hole propagators cannot separate it into two four-point vertices. 
    \item \emph{Vertical irreducibility.} Similarly, a diagram is irreducible in the vertical channel if a horizontal cut through two vertical particle--hole propagators cannot separate it into two four-point vertices. 
\end{itemize}

Let $\Gamma^{\mathrm{ph}}(12;34)$ and $\bar\Gamma^{\mathrm{ph}}(12;34)$ denote the vertices irreducible in the horizontal and vertical channels, respectively. The crossing transformation relates them through
\begin{equation}
    \Gamma^{\mathrm{ph}}(42;31)
    =-\bar\Gamma^{\mathrm{ph}}(12;34).
\end{equation}
The two channels contain exchange-related contributions to the four-point vertex, including those appearing at second order. Starting from an antisymmetrized bare vertex does not, by itself, guarantee that a subsequent resummation preserves crossing symmetry. The horizontal and vertical contributions must also be treated consistently.

Given the vertex irreducible in the horizontal channel, the full vertex satisfies the Bethe--Salpeter equation
\begin{equation}
\begin{aligned}
    \Gamma(12;34)
    ={}&\Gamma^{\mathrm{ph}}(12;34)+\Gamma(12;56)G^{\mathrm{ph}}(56;78)
    \Gamma^{\mathrm{ph}}(78;34),
\end{aligned}
\label{S38}
\end{equation}
where $
    G^{\mathrm{ph}}(56;78)=G(8,6)G(5,7)
$
is the product of the two particle--hole propagators. The diagrammatic form is shown in Fig.~3a of main-text. All diagrams irreducible in the horizontal channel are included in $\Gamma^{\mathrm{ph}}$; the remaining diagrams are generated by connecting vertices through a pair of particle--hole propagators. Rearranging Eq.~\eqref{S38} gives
\begin{align}
    &\Gamma(12;34)
    -\Gamma(12;56)G^{\mathrm{ph}}(56;78)
    \Gamma^{\mathrm{ph}}(78;34)
    =\Gamma^{\mathrm{ph}}(12;34),\\
    &\Gamma(12;56)
    \bigl[\delta_{35}\delta_{46}
    -G^{\mathrm{ph}}(56;78)\Gamma^{\mathrm{ph}}(78;34)\bigr]
    =\Gamma^{\mathrm{ph}}(12;34).
\end{align}
In matrix notation, the solution is
\begin{equation}
    \Gamma=\Gamma^{\mathrm{ph}}
    \bigl[I-G^{\mathrm{ph}}\Gamma^{\mathrm{ph}}\bigr]^{-1}. 
\end{equation}

The preceding equation determines the full vertex once $\Gamma^{\mathrm{ph}}$ is known. To construct this irreducible vertex, let $\Lambda^{\mathrm{irr}}$ denote the contribution that is irreducible in both particle--hole channels. The horizontal-channel irreducible vertex can then be expressed as
\begin{equation}
\begin{aligned}
    \Gamma^{\mathrm{ph}}(12;34)
    ={}&\Lambda^{\mathrm{irr}}(12;34)-\Gamma(15;36)G^{\mathrm{ph}}(56;78)
    \bar\Gamma^{\mathrm{ph}}(72;84).
\end{aligned}
\end{equation}
The corresponding diagram is shown in Fig.~3b of maintext. The minus sign accounts for the fermionic sign of the crossed-channel contribution in the adopted convention.

Using crossing symmetry to relate the two channels yields
\begin{equation}
\begin{aligned}
    \Gamma^{\mathrm{ph}}(12;34)
    ={}&\Lambda^{\mathrm{irr}}(12;34)-\Gamma(13;56)G^{\mathrm{ph}}(56;78)
    \Gamma^{\mathrm{ph}}(78;24).
\end{aligned}
\label{196}
\end{equation}
Substituting the Bethe--Salpeter solution for $\Gamma$, we obtain
\begin{equation}
\begin{aligned}
    \Gamma^{\mathrm{ph}}(12;34)
    & =\Lambda^{\mathrm{irr}}(12;34)-\Phi(13;24),\quad
    \Phi
     =\Gamma^{\mathrm{ph}}
    \bigl[I-G^{\mathrm{ph}}\Gamma^{\mathrm{ph}}\bigr]^{-1}
    G^{\mathrm{ph}}\Gamma^{\mathrm{ph}}.
\end{aligned}
\label{S44}
\end{equation}
This rewriting does not introduce an additional approximation for $\Gamma^{\mathrm{ph}}$. Rather, it provides a self-consistency condition relating the two particle--hole channels. Combining Eqs.~\eqref{S38} and \eqref{S44}, we can express the full vertex as
\begin{equation}
    \Gamma(12;34)
    =\Lambda^{\mathrm{irr}}(12;34)
    -\Phi(13;24)+\Phi(12;34).
    \label{S45}
\end{equation}
This form displays the direct- and crossed-channel contributions explicitly. For a specified $\Lambda^{\mathrm{irr}}$, one first solves Eq.~\eqref{S44} self-consistently for $\Gamma^{\mathrm{ph}}$ and then obtains the full vertex from Eq.~\eqref{S45}.

Preserving the crossing structure in Eq.~\eqref{S45} requires $\Gamma^{\mathrm{ph}}$ to satisfy Eq.~\eqref{S44}; it cannot be approximated independently of this consistency condition. The approximation is specified through $\Lambda^{\mathrm{irr}}$. A simple choice is the Hartree--Fock Coulomb kernel. The principal numerical challenge is the memory required to represent the momentum- and index-dependent vertex $\Gamma^{\mathrm{ph}}$.

{\it Numerics.--}
The numerical problem begins with the Eq.~\ref{S44}. 
With the operator ordering $c^\dagger(1)c(2)c^\dagger(4)c(3)$, the correspondence with the earlier notation $\Gamma^Q_{uv,ij}({\bf q,p})$ is
\begin{equation}
    k_1=q-Q,\qquad k_2=q,\qquad
    k_3=p-Q,\qquad k_4=p.
\end{equation}
Suppressing the band indices for clarity, we write
$\Gamma^{\mathrm{ph}}(q,p;Q)\equiv\Gamma^{\mathrm{ph}}(12;34)$
with this momentum assignment. In the single-channel ladder equation, the transfer momentum
$Q=k_2-k_1=k_4-k_3$ is conserved, so each value of $Q$ can be treated independently. In the crossed contribution, however, the transfer is
$Q'=k_3-k_1=p-q$. Thus, although each matrix operation can be performed at fixed transfer momentum, the crossing relation couples different transfer sectors in the self-consistency equation.

To evaluate $\Phi$ without explicitly forming an inverse, we use the matrix identity
\begin{equation}
    \Gamma^{\mathrm{ph}}
    \bigl[I-G^{\mathrm{ph}}\Gamma^{\mathrm{ph}}\bigr]^{-1}
    =\bigl[I-\Gamma^{\mathrm{ph}}G^{\mathrm{ph}}\bigr]^{-1}
    \Gamma^{\mathrm{ph}},
\end{equation}
which follows from
$
    \bigl[I-\Gamma^{\mathrm{ph}}G^{\mathrm{ph}}\bigr]
    \Gamma^{\mathrm{ph}}
    =\Gamma^{\mathrm{ph}}
    \bigl[I-G^{\mathrm{ph}}\Gamma^{\mathrm{ph}}\bigr].
$
Consequently,
\begin{equation}
\begin{aligned}
    \Phi
    &=\bigl[I-\Gamma^{\mathrm{ph}}G^{\mathrm{ph}}\bigr]^{-1}    \Gamma^{\mathrm{ph}}G^{\mathrm{ph}}\Gamma^{\mathrm{ph}},\quad
    \bigl[I-\Gamma^{\mathrm{ph}}G^{\mathrm{ph}}\bigr]\Phi =\Gamma^{\mathrm{ph}}G^{\mathrm{ph}}\Gamma^{\mathrm{ph}}.
\end{aligned}
\label{203}
\end{equation}
Equation~\eqref{203} is a linear matrix equation for $\Phi$ at fixed $Q$, given $\Gamma^{\mathrm{ph}}$. 

One possible approach is Newton's method. Starting from an initial estimate $\Gamma_n^{\mathrm{ph}}$, we evaluate $\Phi_n$ using Eq.~\eqref{203} and compute the residual
\begin{equation}
    R[\Gamma_n^{\mathrm{ph}}]
    =\Gamma_n^{\mathrm{ph}}-\Lambda^{\mathrm{irr}}
    -\mathcal C\Phi_n,
\end{equation}
where $\mathcal C$ denotes the crossing transformation. If the residual exceeds the chosen tolerance, we update the vertex according to
\begin{equation}
    \Gamma_{n+1}^{\mathrm{ph}}
    =\Gamma_n^{\mathrm{ph}}+\alpha H.
    \label{205}
\end{equation}
Here $\alpha$ is the step size, and $H$ is obtained from the global Newton linear system. Specifically, $H$ is chosen so that the linearized residual vanishes:
\begin{equation}
    R[\Gamma_n^{\mathrm{ph}}+H]
    \simeq R[\Gamma_n^{\mathrm{ph}}]
    +J[\Gamma_n^{\mathrm{ph}}]H=0,
\end{equation}
where $J$ is the Jacobian of the residual. After each update, we recompute $\Phi_{n+1}$ and evaluate the new residual. Once $\Gamma^{\mathrm{ph}}$ has converged to within the chosen tolerance, the full vertex follows from Eq.~\ref{S45}.

\section{Superconductivity from a Crossing-Symmetric Parquet Vertex}

We now use the crossing-symmetric, particle--hole-renormalized vertex
$\Gamma$ to formulate the superconducting pairing problem. If all microscopic degrees of freedom are retained, the simplest
approximation is to take $\Lambda^{\mathrm{irr}}$ to be the bare Coulomb
vertex, including both direct and exchange terms. Such a calculation is,
however, computationally demanding. As a practical alternative, we first
screen the Coulomb interaction within the random-phase approximation (RPA),
excluding particle--hole processes in which both internal states belong to
the active band. We denote the resulting interaction by
$V^c_{\mathrm{RPA}}$ and construct $\Lambda_c^{\mathrm{irr}}$ from its direct
and exchange contributions. This screened vertex then serves as the input
to the parquet equations for $\Gamma$. 

For either choice of input interaction, the vertex
$\Gamma(k,k+q;-k-q+P_0,-k+P_0)$ describes a pairing process with total pair momentum
$P_0$. Determining the favored pairing state
therefore requires retaining $P_0$ as a general parameter and comparing
the corresponding free energies. Below we consider $P_0=0$. We define the equal-time anomalous correlation function as
\begin{equation}
d(\mathbf{k})\equiv
\left\langle
\psi(\mathbf{k},t)\psi(-\mathbf{k},t)
\right\rangle\neq 0.
\end{equation}
Here $\psi$ denotes the active band operator. We seek a static mean-field solution for which equal-time correlation
functions are independent of time. Suppressing the active-band indices,
we write the pairing interaction and its mean-field decoupling as
\begin{align}
V_{\mathrm{MF}}
&=-\frac{1}{4}\sum_{\mathbf{k},\mathbf{q}}
\Gamma(\mathbf{k},-\mathbf{k};\mathbf{q})\,
\psi^\dagger(\mathbf{k})\psi^\dagger(-\mathbf{k})
d(\mathbf{k}+\mathbf{q})
+\mathrm{h.c.}+\mathrm{constant}.
\label{S53}
\end{align}
Here the three-argument notation for the vertex is defined by $
\Gamma(\mathbf{k},-\mathbf{k};\mathbf{q})
\equiv\Gamma({\bf k,k+q;-k-q,-k}).
$
Defining the gap function as
\begin{equation}
\Delta(\mathbf{k})=\frac{1}{2}\sum_{\mathbf{q}}
\Gamma(\mathbf{k},-\mathbf{k};\mathbf{q})\,
d(\mathbf{k}+\mathbf{q}),
\end{equation}
we obtain the Bogoliubov--de Gennes (BdG) Hamiltonian
\begin{equation}
H_{\mathrm{MF}}
=\frac{1}{2}\sum_{\mathbf{k}}
\begin{pmatrix}
c^\dagger_{\mathbf{k}}&c_{-\mathbf{k}}
\end{pmatrix}
\begin{pmatrix}
\xi(\mathbf{k})&-\Delta(\mathbf{k})\\
-\Delta^\dagger(\mathbf{k})&-\xi(-\mathbf{k})
\end{pmatrix}
\begin{pmatrix}
c_{\mathbf{k}}\\c^\dagger_{-\mathbf{k}}
\end{pmatrix}.
\label{59}
\end{equation}
Here $\xi(\mathbf{k})$ is the active-band energy measured from the chemical
potential. The prefactor $1/2$ compensates for double counting in the
Nambu representation. For the isospin-polarized,
single-band state considered here, fermionic antisymmetry requires
$\Delta(-\mathbf{k})=-\Delta(\mathbf{k})$.

The quadratic kernel in the imaginary-time action is
$-i\omega_n+H_{\mathrm{MF}}(\mathbf{k})$, where $H_{\mathrm{MF}}(\mathbf{k})$ denotes the BdG
matrix in Eq.~\eqref{59}. The Nambu Green's function
therefore takes the following form in momentum--frequency space:
\begin{equation}
G(i\omega_n,\mathbf{k})
=-\left[-i\omega_n+H_{\mathrm{MF}}(\mathbf{k})\right]^{-1}
=\begin{pmatrix}
i\omega_n-\xi(\mathbf{k})&\Delta(\mathbf{k})\\
\Delta^\dagger(\mathbf{k})&i\omega_n+\xi(-\mathbf{k})
\end{pmatrix}^{-1}.
\end{equation}
The overall factor of $1/2$ in the Nambu Hamiltonian does not enter this
matrix inverse: it compensates for the redundant summation over
$\mathbf{k}$ and $-\mathbf{k}$, whereas the dynamics at fixed momentum
are governed by $H_{\mathrm{MF}}(\mathbf{k})$. This can also be seen from the
equations of motion or by restricting the momentum sum to one
representative of each pair $\{\mathbf{k},-\mathbf{k}\}$.

The normal and anomalous components are
\begin{align}
G_{11}(i\omega_n,\mathbf{k})
&=\frac{i\omega_n+\xi(-\mathbf{k})}
{\displaystyle\prod_{\alpha=1,2}
\left(i\omega_n-E_\alpha\right)},\quad
G_{12}(i\omega_n,\mathbf{k})
=-\frac{\Delta(\mathbf{k})}
{\displaystyle\prod_{\alpha=1,2}
\left(i\omega_n-E_\alpha\right)}.
\end{align}
Here $G_{12}$ is the frequency space of the anomalous correlation
$-\langle\psi_{\mathbf{k}}(\tau)\psi_{-\mathbf{k}}(0)\rangle$.
The gap equation is consequently
\begin{equation}
\Delta(\mathbf{k})
=\frac{1}{2}T\sum_{n,\mathbf{q}}
\Gamma(\mathbf{k},-\mathbf{k};\mathbf{q}-\mathbf{k})
\frac{\Delta(\mathbf{q})}
{\displaystyle\prod_{\alpha=1,2}
\left(i\omega_n-E_\alpha\right)},
\end{equation}
where the quasiparticle energies in the denominator are evaluated at
$\mathbf{q}$. Performing the Matsubara sum gives
\begin{equation}
T\sum_n
\frac{1}{(i\omega_n-E_1)(i\omega_n-E_2)}
=\frac{n_F(E_1)-n_F(E_2)}{E_1-E_2}.
\end{equation}
We thus obtain the self-consistent gap equation
\begin{equation}
\Delta(\mathbf{k})
=\frac{1}{2}\sum_{\mathbf{q}}
\Gamma(\mathbf{k},-\mathbf{k};\mathbf{q}-\mathbf{k})\,
\Delta(\mathbf{q})\,
\frac{n_F(E_1)-n_F(E_2)}{E_1-E_2}.
\end{equation}
The energies $E_1$ and $E_2$ are the eigenvalues of the BdG matrix at
$\mathbf{q}$. To determine the superconducting state at fixed particle density, we
solve the gap equation together with the number equation,
\begin{equation}
n=\int\frac{d^2k}{(2\pi)^2}\,n(\mathbf{k}),
\qquad
n(\mathbf{k})=\sum_i
\Psi_{1i}(\mathbf{k})\,
\frac{1}{e^{\beta E_i(\mathbf{k})}+1}\,
\Psi^*_{1i}(\mathbf{k}).
\end{equation}
Here $\Psi_{1i}(\mathbf{k})$ is the first component of the normalized BdG
eigenvector associated with $E_i(\mathbf{k})$.